**Estimation of Protein-Ligand Unbinding Kinetics Using Non-Equilibrium Targeted Molecular Dynamics Simulations**

Steffen Wolf*[†,‡,#], Marta Amaral[§,||,∪,#], Maryse Lowinski[⊥], Francois Vallée[⊥], Djordje Musil[||], Jörn Güldenhaupt[†], Matthias K. Dreyer[∠], Jörg Bomke[∩], Matthias Frech[||], Jürgen Schlitter[†], Klaus Gerwert[†]

---

† Department of Biophysics, Ruhr-University Bochum, 44780 Bochum, Germany

‡ Institute of Physics, Albert-Ludwigs-University Freiburg, 79104 Freiburg, Germany

§ Instituto de Biologia Experimental e Tecnológica, 2780-157 Oeiras, Portugal

|| Molecular Interactions and Biophysics, Merck KGaA, 64293 Darmstadt, Germany

∪ Sanofi-Aventis Deutschland GmbH, Biologics Research / Protein Therapeutics, 65926 Frankfurt am Main, Germany

⊥ Sanofi IDD-BioStructure and Biophysics, 94400 Vitry-sur-Seine, France

∠ Sanofi-Aventis Deutschland GmbH, R&D Integrated Drug Discovery, 65926 Frankfurt am Main, Germany

∩ Molecular Pharmacology, Merck KGaA, 64293 Darmstadt, Germany

# These authors contributed equally.

*steffen.wolf@physik.uni-freiburg.de

**Abstract**

We here report on non-equilibrium targeted Molecular Dynamics simulations as tool for the estimation of protein-ligand unbinding kinetics. Correlating simulations with experimental data from SPR kinetics measurements and X-ray crystallography on two small molecule compound libraries bound to the N-terminal domain of the chaperone Hsp90, we show that the mean non-equilibrium work computed in an ensemble of trajectories of enforced ligand unbinding is a promising predictor for ligand unbinding rates. We furthermore investigate the molecular basis determining unbinding rates within the compound libraries. We propose ligand conformational changes and protein-ligand nonbonded interactions to impact on unbinding rates. Ligands may remain longer at the protein if they exhibit strong electrostatic and/or van der Waals interactions with the target. In the case of ligands with rigid chemical scaffold that exhibit longer residence times however, transient electrostatic interactions with the protein appear to facilitate unbinding. Our results imply that understanding the unbinding pathway and the protein-ligand interactions along this path is crucial for the prediction of small molecule ligands with defined unbinding kinetics.

**Introduction**

While rational drug design traditionally focuses on the optimization of binding affinity of compounds to target proteins, optimization of target binding kinetics is emerging as a new paradigm in drug discovery.[1-7] Often, drugs with optimized binding kinetics exhibit better efficacy profiles and reduced off-target toxicity,[1,8] and thus are more likely to pass later clinical phases.[9] However, while the prerequisites for the rational design of high affinity drugs are well investigated,[10] the rational optimization of kinetic parameters of small molecules is in its early stages.[11,12] Molecular determinants believed to be important in the modulation of binding kinetics include ligand molecular size, hydrophobic effects, electrostatic interactions, and conformational fluctuations.[4,11] Recent reports further highlight the importance of protein-bound water molecules[12] and of protein internal electrostatic interactions.[13] However, the exact contribution and extent of each of these properties still needs to be further elucidated.

In order to gain a systematic understanding of the impact of different molecular discriminants on binding kinetics, and thus help to establish a knowledge basis necessary for rational design of compounds with desired kinetics, we performed a combined experimental and theoretical analysis on the dynamics of unbinding of two series of compounds with different chemical scaffolds (see Figure 1) bound to the ATP-binding N-terminal domain of the chaperone heat shock protein 90 (Hsp90, Figure 1C),[14-16] which is a well-known target for anti-cancer drugs.[14,17-19] Based on data shared within the Kinetics for Drug Discovery consortium (K4DD, www.k4dd.eu)[7,20,21] and preexisting data sets,[19,22,23] we included a total of 26 compounds in the present analysis, which are listed in Table 1. Additionally, we determined by X-Ray crystallography the structures of one further protein-ligand complex (see Tables 1 and S2), and measured ligand binding kinetics and affinities of three further compounds via surface plasmon resonance (SPR). In detail, we investigated fourteen compounds with resorcinol backbone (compounds **1a-1n,** see Figure 1; amongst them the Hsp90 inhibitor Ganetespib[24] **1c**), eleven compounds with N-heterocycle functionalities[19] (compounds **2a-2k**), and the macrocyclic lactam Hsp90 inhibitor 17-DMAG[18] **17**. Figure 1C displays an overview of the N-terminal domain of Hsp90 with bound compound **1f**. The binding site is located close to the protein surface, and exhibits two different conformations of the adjacent amino acids 102-114. These residues either form a helix conformation (helix 3) or a loop conformation, which was proposed to affect unbinding kinetics.[7]

**Table 1**. List of compounds, dynamic properties and protein conformations of investigated compounds. Error bars denote the standard error of the mean (SEM) for N=2-4 measurements. n.d.: not determined.

| compound | Ref. for kinetics and affinity | $k_{off}$ / $s^{-1}$ | $K_D$ / $M^{-1}$ | $k_{on}$ / $M^{-1}$ $s^{-1}$ | helix 3 conf. | PDB ID with ref. |
|---|---|---|---|---|---|---|
| **1a** | 2 | <1.00E-04 | <1.00E-09 | (n.d.) | loop | 5NYI[20] |
| **17** (17-DMAG) | 1 | 3.00E-04 | 4.57E-09 | (n.d.) | loop (out) | 1OSF[22] |
| **1b** | 1 | 3.30E-04<br>±2.1E-05 | 4.60E+05<br>±4.0E-11 | 2.15E+05<br>±5.40E+04 | helix | 5J20[7] |
| **1c** (Ganetespib) | 1 | 5.70E-04 | 1.00E-09 | (n.d.) | loop | 3TUH[23] |
| **1d** | 1 | 1.70E-03<br>±4.6E-04 | 2.30E-08<br>±4.4E-09 | 7.00E+04<br>±7.50E+03 | helix | 5J9X[7] |
| **1e** | 1 | 1.79E-03<br>±4.7E-06 | 3.81E-09<br>±3.5E-10 | 4.72E+05<br>±4.10E+04 | helix | 5J86[7] |
| **1f** | 1 | 4.20E-03<br>±5.2E-04 | 2.40E-08<br>±1.0E-09 | 1.80E+05<br>±2.50E+04 | helix | Modelled<br>from M5 |
| **1g** | 1 | 6.40E-03<br>±4.3E-04 | 8.70E-08<br>±2.1E-09 | 7.70E+04<br>±1.20E+04 | helix | 5J27[20] |
| **1h** | 1 | 1.40E-02<br>±2.2E-03 | 2.66E-08<br>±2.6E-09 | 5.20E+05<br>±1.30E+05 | loop | 5J2X[7] |
| **1i** | 1 | 1.40E-02<br>±1.5E-03 | 4.30E-07<br>±6.1E-08 | 3.30E+04<br>±1.30E+03 | helix | 5J86[20] |
| **1j** | 1 | 3.38E-02<br>±1.13E-03 | 7.11E-08<br>±4.327E-9 | 4.79E+05<br>±1.65E+04 | loop | 6FCJ[21] |
| **1k** | 2 | 6.34E-02<br>±3.5E-03 | 5.14E-07<br>±6.0E-09 | 1.23E+05<br>±5.23E+03 | loop | 6ELO[20] |
| **1l** | 2 | 1.74E-01<br>±2.2E-02 | 2.36E-07<br>±1.9E-08 | 7.42E+05<br>±1.53E+05 | helix | 6ELP[20] |
| **1m** | 1 | 2.10E-01<br>±3.3E-02 | 1.80E-07<br>±1.2E-08 | 1.20E+06<br>±2.10E+05 | loop | 5J64[7] |
| **1n** | 2 | 2.54E-01<br>±1.8E-02 | 9.00E-07<br>±1.7E-08 | 2.80E+05<br>±1.47E+04 | loop | 6ELN[20] |
| **2a** | (here) | 7.10E-05<br>±4.0E-06 | 7.74E-09<br>±4.0E-10 | (n.d.) | helix | 2YKC[19] |
| **2b** | 2 | 1.36E-04<br>±3.8E-06 | 8.48E-09<br>±6.9E10 | 1.62E+04<br>±1.78E+03 | helix | 5LQ9[20] |
| **2c** | 2 | 1.89E-04<br>±7.1E-05 | 4.66E-08<br>±2.5E-08 | 4.77E+03<br>±1.35E+03 | helix | 5LR7[20] |
| **2d** | (here) | 1.94E-04<br>±9.0E-06 | 1.01E-08<br>±5.0E-10 | (n.d.) | helix | 5LRL (here) |
| **2e** | 2 | 2.78E-04<br>±4.65E-06 | 1.72E-07<br>±1.2E-07 | 3.06E+03<br>±2.09E+03 | helix | 5LRZ[20] |
| **2f** | 2 | 2.85E-04<br>±4.9E-05 | 3.61E-08<br>±5.7E-09 | 7.77E+03<br>±1.48E+02 | helix | 2YKI[19] |
| **2g** | 2 | 4.85E-04<br>±1.39E-04 | 3.95E-09<br>±1.7E-09 | 1.33E+05<br>±2.37E+04 | helix | 5LS1[20] |
| **2h** | 2 | 7.65E-04<br>±5.0E-05 | 2.40E-10<br>±8.9E-11 | 3.58E+06<br>±1.11E+06 | helix | 5T21[20] |
| **2i** | 2 | 9.89E-04<br>±1.3E-04 | 9.50E-08<br>±4.5E-09 | 1.04E+04<br>±9.04E+02 | helix | 2YKJ[19] |
| **2j** | (here) | 3.697E-03<br>±1.5E-05 | 3.285E-8<br>±5.61E-9 | 1.17E+05<br>±1.49E4 | helix | 5LO1[25] |
| **2k** | 2 | 2.56E-02<br>±1.4E-02 | 2.47E-08<br>±6.5E-09 | 1.26E+06<br>±8.78E+05 | helix | 5LR1[20] |

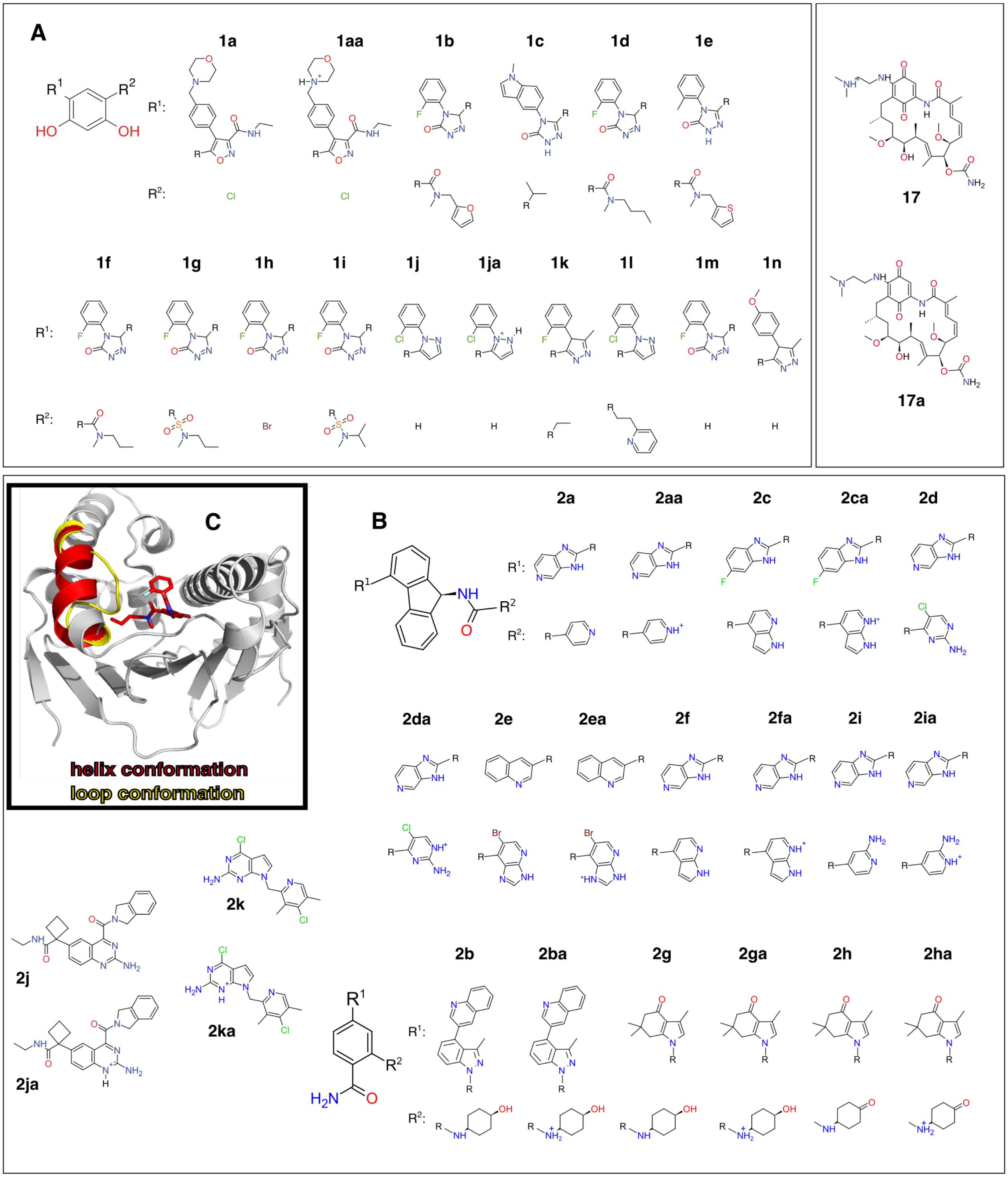

**Figure 1.** Structure of the N-terminal domain of Hsp90 and of investigated ligand scaffolds. Compound names with two letters denominate alternative protonation states. A: resorcinol compounds (**1a-1n**) and additional compound **17** (17-DMAG). B: N-heterocycle series (**2a-2k**) with fluorenamide and benzamide scaffolds and additional compounds **2j** and **2k**. C: overview of the N-terminal domain of HSP90 in complex with compound **1f**. Protein in cartoon, compound **1f** in sticks, helix 3 in red, alternate loop conformation in yellow.

To assess the molecular mechanisms of unbinding in Hsp90, we performed non-equilibrium targeted molecular dynamics (TMD) simulations.[26-29] This method uses a constant velocity constraint as an additional force $f_c$ in the simulations to push the ligand out of the binding site. $f_c$ is calculated via a Lagrange multiplier with regards to the ligand's center of mass (COM) and is updated each time step to move the ligand to a position that is in agreement with the preset constant velocity. The constraint force is applied in such a way that the distance between anchor group COM and ligand COM is increased with the preset constant velocity (see Fig. S1). The distance vector acts like a radial vector in a spherical coordinate system, while the ligand is free to move and change direction on the surface of the sphere.[26] This leaves the ligand the freedom to perform diffusion perpendicular to the distance vector, conformational changes and rotations. The ligand thus has the possibility to probe different unbinding pathways, although this choice is limited by the ratio between constraint velocity and diffusion on the sphere. We focus our analysis on the contributions to unbinding kinetics, as unbinding events are easier to calculate than binding events.[30] As we almost exclusively use protein/ligand crystal structures with positions of protein-internal water molecules being resolved, we have an excellent structural basis for carrying out such simulations. As the simulations are carried out under non-equilibrium conditions, i.e., non-stationary with a finite time and velocity, we do not obtain the free energy along the pulling coordinate, but a non-equilibrium work $\langle W \rangle$. According to the second law of thermodynamics, $\Delta G \leq W$ due to W containing irreversible work caused by dissipation, i.e., friction effects. We find that this non-equilibrium variable is a better predictor for unbinding kinetics than free energy profiles derived from stationary free energy calculations.

## Methods

**Chemistry and analytical data of 1j and 2j.** Information on the synthesis of chemical compounds **1j** is provided in refs. 7,20, on **2j** in ref. 25 and their analytical data in Table S1. LC/MS spectra of the products were recorded on an Agilent 1100 HPLC system (1100 high pressure gradient pump, 1100 diode array detector) interfaced to an Agilent 1100 mass spectrometer detector using a Chromolith SpeedROD RP 18e50-4.6 column. Polar gradient: Water (0.05% HCOOH) - acetonitrile (0.04% HCOOH) were used as eluent in mixtures as follows: 0 min, 4% ACN; 2.8 min, 100% ACN; 3.3 min, 100% ACN; Gradient: 5.5 min; Flow-rate: 2.4 ml/min; UV detection: 220 nm. $^{1}$H NMR spectra were recorded at 300 K unless otherwise specified using a Bruker Avance DPX 300, AV 400, DPX 500 spectrometer (TMS as an internal standard). 1H NMR chemical shifts are reported in parts per million (ppm). 1 H NMR data is reported as chemical shift (dH), relative integral, multiplicity (s = singlet, d = doublet, t = triplet, q = quartet, dd = doublet of doublets, ddd = doublet of doublet of doublets, dt = doublet of triplets, td = triplet of doublets, tt = triplet of triplets, qd = quartet of doublets) and coupling constant (J Hz). Both compounds have a purity ≥95%.

Information on the synthesis of chemical compounds **1a**-**1i** and **1k**-**1n** can be found in patent WO2006087077 and in refs. 7,20,25; for compounds **2a**-**2f**, and **2i** in published patent applications WO2010106290, WO2006123061, WO2008049994 and ref. 19, for compounds **2g** and **2h** in patent WO2006091963, and **2k** in patent WO2005028434.

**Crystallization and Structure Determination for compound 2d:** A hexa-histidine tagged N-terminal fragment of Hsp90 (18-223) (NP_005339) was expressed and purified as described in ref. 20. Crystallization conditions are also described in ref. 20. Datasets were collected in-house on a Rigaku HF-007 rotating anode generator and a MAR CCD detector and in the synchrotron.

Diffraction data were processed with either XDS[31] or MOSFLM.[32] The structures were solved by the molecular replacement method using one set of coordinates of N-HSP90 available in the Protein Data Bank (PDB code: 1YER). The structures were refined using either CNX7,[33] REFMAC5[34] or BUSTER8 program packages,[35] ligands were placed manually, and the structural models were manually rebuilt using either TURBO-FRODO (www.afmb.univ-mrs.fr/-TURBO) or COOT9[36]. Final validation checks were performed using MOLPROBITY.[37]

**Surface Plasmon Resonance (SPR) of compounds 2a, 2d and 2j:** SPR measurements were performed on a Biacore 4000 instrument from GE Healthcare as previously described in refs. 7,20. Briefly, recombinant N-HSP90 with 17-desmethoxy-17-N,N-dimethylaminoethylamino-geldanamycin (17-DMAG, Merck Millipore) was immobilized on a Biacore CM5 chip at 25°C at a flow rate of 10 μL/min using amine coupling at pH 4.50 according to Biacore's standard protocol. HBS-N (10 mM Hepes pH 7.40, 0.15 M NaCl) served as the running buffer during immobilization and all SPR binding kinetics measurements assays were performed in 20 mM HEPES pH 7.50, 150 mM NaCl, 0.05% Tween 20, 1 mM DTT, 0.1 mM EDTA, 2% DMSO. Data sets were processed and analyzed using the Biacore 4000 Evaluation software, version 1.1. Solvent corrected and double-referenced association and dissociation phase data were fitted to a simple 1:1 interaction model with mass transport limitations.

**Simulation setup:** TMD calculations were performed with Gromacs v4.6.5 (ref. 38) using the AMBER99SB forcefield[39,40] for protein and ions, and the TIP3P water model.[41] Crystal structures for compounds **17** and **1c** were taken from PDB IDs 1OSF[22] and 3TUH,[23] respectively. Structures of compounds **2a, 2f** and **2i** were taken from PDB IDs 2YKC, 2YKI and 2YKJ.[19] Due to their high similarity, the structure of compound **1f** was modeled based on the **1d** protein-ligand complex by removing a single terminal methyl group of the respective butenyl side chain. The initial structure of compound **2d** was taken from the structure published herein. Crystal structures of all other compounds were determined within the Kinetics for Drug Discovery consortium and are published in refs. 7,20,21 (see Table 1). Ligand parameters were created with antechamber[42] and acpype[43] using GAFF parameters[44] and AM1-BCC charges.[45,46] Protein/ligand crystal structures together with present crystal water molecules were centered in a cubic box with 7 nm side length, missing protons added, protonated, solvated, and sodium ions added to ensure a charge neutral simulation box. Protonation states of amino acids were determined by propka.[47] Protonation states of other compounds were determined based on $pK_a$ values predicted using Chemicalize[48] developed by ChemAxon (http://www.chemaxon.com). Besides prediction of protonation states for pH = 7.5, we checked at which pH the ligand exhibits a 10% population of alternative protonated states.

**TMD calculations:** Simulations were carried out with PME[49] for electrostatics (minimal real space cut-off of 1 nm) and a van der Waals cut-off of 1 nm. Hydrogen atom bonds were constrained via the LINCS[50] algorithm. The prepared systems were first minimized with the conjugate gradient method, and subjected to a short equilibration runs in the NPT ensemble at 300 K and 1 bar, using the Berendsen thermostat and barostat,[51] with an integration step size of 2 fs and a trajectory length of 100 ps. For each ligand, 30 statistically independent equilibration runs were performed, in which differed velocity distributions were attributed to the minimized systems to generate an initial equilibrium state ensemble. Non-equilibrium TMD calculations using the Gromacs PULL code in constraint mode were then carried out by continuing the 30 independent equilibration runs for 200 ps in the NPT ensemble at 300 K and 1 bar, using the Nosé-Hoover thermostat[52,53] and Parrinello-Rahman barostat,[54] with a fixed constraint velocity of 0.01 nm/ps and an integration step size of 1 fs. Constraint pseudoforces were written out for each time step. The 1$^{st}$ reference group for COM pulling along path 1 consisted of all C(alpha) atoms of the beta-sheet forming the ligand binding site (see Fig. S1) and of all C(alpha) atoms of helix 1 for path 2, the 2$^{nd}$ group was formed by the ligand heavy atoms. Integrating $f_c$ along the pathway as

$$W(x) = \int_{x_0}^{x} f_c(x')\, dx' \tag{1}$$

yields the non-equilibrium work $W$ performed to remove the ligand. In our simulations, we obtain a resolution of 10 fm along *x,* and calculate $W$ via trapezoidal numeric integration. Trajectory evaluation was then carried out with Gromacs tools, and data evaluation in Python using numpy and scipy libraries[55] and Jupyter notebooks. [56]

Stationary thermodynamic integration[57] simulations were performed by extracting 21 equidistant snapshots from a random non-equilibrium simulations, and carrying out equilibration simulations of 10 ns trajectory length with them, setting the constraint velocity to zero (for a detailed explanation see ref. 28). Mean constraint pseudoforces $\langle f_c \rangle$ were calculated from the last 2.5 ns of these simulations. Free energy profiles as given in Fig. S2A were then calculated by integrating the mean forces along the distance from the binding site *x* as

$$\Delta G(x) = \int_{x_0}^{x} \langle f_c(x') \rangle dx' \approx \sum_{i}^{N_x} \langle f_{c,i} \rangle \, \Delta x \tag{2}$$

with $N_x$ being the number of steps between $x_0$ and x, and $\Delta x$ the distance between snapshots.

**Error calculation:** For experimental results, we calculated errors as standard error of the mean (SEM) $\Delta x = \sigma/\sqrt{N}$ with standard deviation σ and number of experiments *N.* To calculate the errors of theoretical results, we used random sampling bootstrapping with replacement as implemented in the Python scikit-learn library,[58] using 10000 iterations. To keep comparability with the experimental SEM,[59] the displayed error bars represent the 1σ confidence level.

**Table 2**. Computational results for mean work <W>, predicted $pK_a$ values of investigated compounds, and pH at which the alternate protonation state forms 10% of the ligands ensemble. Dominant form denotes the protonation state at pH = 7.5. Error bars indicate the 1σ confidence interval from bootstrap analysis (see Methods)

| compound | <W> /<br>kJ/mol | $pK_a$ | pH for 10%<br>ensemble presence |
|---|---|---|---|
| **1a**<br>**1aa** | 515.2 ± 14.6<br>527.8 ± 15.3 | 8.0 | 7.1<br>-dominant form- |
| **17** (17-DMAG)<br>**17a** | 558.8 ± 10.2<br>591.4 ± 12.2 | 9.8 | 8.8<br>-dominant form- |
| **1b** | 577.6 ± 10.8 | | |
| **1c** (Ganetespib) | 473.0 ± 10.4 | | |
| **1d** | 511.9 ± 17.8 | | |
| **1e** | 559.2 ± 13.3 | | |
| **1f** | 534.8 ± 10.8 | | |
| **1g** | 529.0 ± 18.0 | | |
| **1h** | 444.4 ± 16.3 | | |
| **1i** | 486.9 ± 13.6 | | |
| **1j**<br>**1ja** | 344.4 ± 12.5<br>370.8 ± 15.3 | 4.2 | -dominant form-<br>5.2 |
| **1k** | 444.3 ± 13.6 | | |
| **1l** | 463.2 ± 9.5 | | |
| **1m** | 416.3 ± 11.7 | | |
| **1n** | 440.4 ± 9.5 | | |
| **2a**<br>**2aa** | 549.6 ± 18.7<br>507.4 ± 15.0 | 3.5 | -dominant form-<br>4.2 |
| **2b**<br>**2ba** | 610.7 ± 16.6<br>683.5 ± 24.3 | 2.9 | -dominant form-<br>3.6 |
| **2c**<br>**2ca** | 652.1 ± 21.8<br>541.6 ± 12.3 | 1.8 | -dominant form-<br>2.8 |
| **2d**<br>**2da** | 560.3 ± 13.1<br>633.3 ± 18.4 | 1.6 | -dominant form-<br>2.3 |
| **2e**<br>**2ea** | 518.0 ± 12.6<br>623.1 ± 18.5 | 2.6 | -dominant form-<br>3.5 |
| **2f**<br>**2fa** | 616.4 ± 17.9<br>541.9 ± 14.7 | 1.9 | -dominant form-<br>2.6 |
| **2g**<br>**2ga** | 448.4 ± 10.0<br>478.0 ± 13.7 | 3.1 | -dominant form-<br>4.0 |
| **2h**<br>**2ha** | 442.5 ± 9.3<br>483.5 ± 12.7 | 3.1 | -dominant form-<br>4.0 |
| **2i**<br>**2ia** | 555.4 ± 13.2<br>504.7 ± 12.6 | 4.8 | -dominant form-<br>5.9 |
| **2j**<br>**2ja** | 386.7 ± 11.1<br>343.1 ± 8.6 | 2.5 | -dominant form-<br>3.3 |
| **2k**<br>**2ka** | 415.3 ± 9.8<br>352.6 ± 12.2 | 3.5 | -dominant form-<br>4.2 |

## Results & Discussion

**A linear non-equilibrium energy relationship for unbinding kinetics.** At the beginning of our investigations, we attempted to characterize ligand unbinding kinetics of Hsp90 ligands by determining their free energy profile along the unbinding pathway via standard stationary Thermodynamic Integration[57] (TI) calculations. The most probable unbinding pathway for the ligand appeared to be the passage through an opening between helices 1 and 3 (pathway 1 in Figure S1). Figure S2A displays the resulting free energy surface for compounds **1b**, **1g**, and **1l**, which is in good general agreement with free energy curves for other Hsp90 binding ligands obtained by umbrella sampling.[25] The three investigated compounds exhibit 1-2 free energy barriers between the ligand bound and unbound state. Interpreting the shape and peak height by means of the Eyring equation[60] for rate constants,

$$k_{off} = \kappa \exp(-\beta \Delta G^{\neq}), \qquad (3)$$

with the friction-dependent prefactor $\kappa$, the inverse temperature $\beta = 1/RT$ and the free energy difference between bound state and unbinding transition state $\Delta G^{\neq}$, we find that **1l** effectively does not exhibit a barrier, but a slope between bound and unbound state, and consequently should be the fastest unbinding of the three test compounds. **1b** and **1g** exhibit a comparable transition barrier of ca. 65 kJ/mol, and both a 2nd small barrier at 1.8 nm. **1b** and **1g** should therefore unbind equally fast, which does not agree with the experimental results (see Table 1). Apparently, $\Delta G^{\neq}$ is not sufficient as descriptor for predicting unbinding kinetics, and would require taking into account the prefactor $\kappa$ in Equation (3), which according to Kramers includes friction effects.[61] A further problem we faced when applying stationary TI calculations was the large number of necessary equilibration points along the unbinding pathway that need several nanoseconds of equilibration for reliable determination of the free energy surface,[29] significantly raising the computational cost for investigating a large set of compounds. Furthermore, in our two investigated compound groups, about half of all compounds exhibit two possible protonation states (**1a**, **17**, **1j** and the full series **2**). As an example, the morpholine side chain in **1aa** can exist in a protonated state with a charge of +1 *e* (see Figure 1A), or in a deprotonated state **1a** with a charge of 0 *e*. All ligands in compound group **2** are bound to the protein by a hydrogen bond between nitrogen atoms in aromatic rings ($pK_a$ range of ca. 1.5-5) and Asp93 (see Figure S3), or via highly polarized water molecules mediating this contact.[62] Assigning the correct protonation state for such protein-ligand-water complexes is a challenging task, as the protein environment can significantly alter $pK_a$ values.[62,63] To avoid a bias from wrongly chosen charge states, we needed a method that allowed us to carry out simulations of multiple compounds in 2-3 possible protonation states, with TI calculations simply being too inefficient for this task.

Surprisingly, when we looked at the mean non-equilibrium work profiles <W> from simulations necessary to generate start coordinates for TI calculations (see Figure S2B), we found that the difference in <W> during simulations qualitatively matches the order of unbinding constants of compounds **1b**, **1g**, and **1l**. Apparently, taking account of dissipation effects improves the quality of the prediction of unbinding constants over TI calculations. Differences in <W> between compounds (Figure S2C) appear at positions where the ΔG curve from TI exhibits local maxima. Furthermore, <W> converges rapidly within already N=30 independent trajectories (see Figure S4). We thus evaluated a possible correlation between non-equilibrium TMD work <W> and experimentally determined $k_{off}$ constants using the full investigated compound set comprising all possible protonation states, as displayed in Figure S5A. As in the case with compounds **1b**, **1g**, and **1l**, we observe a qualitative agreement between <W>(TMD) and $k_{off}$, that appears to follow a linear dependency, with ligands requiring a large <W> being slowly unbinding compounds. Such a linear dependence can be expected for equilibrium $\Delta G^{\neq}$ in form of a linear free energy relationship,[64] but is surprisingly present in our non-equilibrium work, as well. The Jarzynski equality[65] connects these two quantities as

$$\exp(-\beta\Delta \mathrm{G}) = \langle \exp(-\beta W)\rangle, \tag{4}$$

where <.> denotes the expectation value of non-equilibrium trajectories based on Boltzmann distribution weights of their initial equilibrium start configurations. [66,67] Following the "fast growth" approach of Hendrix and Jarzynski,[68] the expectation value can be calculated as arithmetic mean

$$\langle \exp\left(-\beta W\right)\rangle \approx \frac{1}{N}\sum_i^N \exp\left(-\beta W_i\right) \tag{5}$$

of the individual values $W_i$ from $N$ non-equilibrium simulations starting from a representative sample of Boltzmann-distributed equilibrium structures. The equality (4) can be recast by a cumulant expansion into

$$\Delta G = \langle W\rangle - W_{diss} \tag{6}$$

with dissipative work $W_{diss}$, which contains the second and all higher moments. [69] After an increase of <W> at transition states $\Delta G^{\neq}$ (cf. Figure S3B,C), we do not observe system relaxation after crossing over the transition states. We thus postulate that

$$\Delta G^{\neq} \approx \langle W\rangle - W_{diss}. \tag{7}$$

Introducing Equation (7) in (3), we obtain

$$\langle W\rangle = -\beta^{-1}\ln k_{off} + C \tag{8}$$

with $C = \beta^{-1}\ln\kappa + W_{diss}$, which serves as a basis of understanding the apparent linear non-equilibrium energy relationship. $C$ effectively is a function of $\beta$, but in the following is treated as an independent fit factor, as we otherwise encountered instabilities in non-linear curve fitting. In the following, we approximate $C$ to be constant, which is only valid in the case that the friction during unbinding is the same for all ligands.

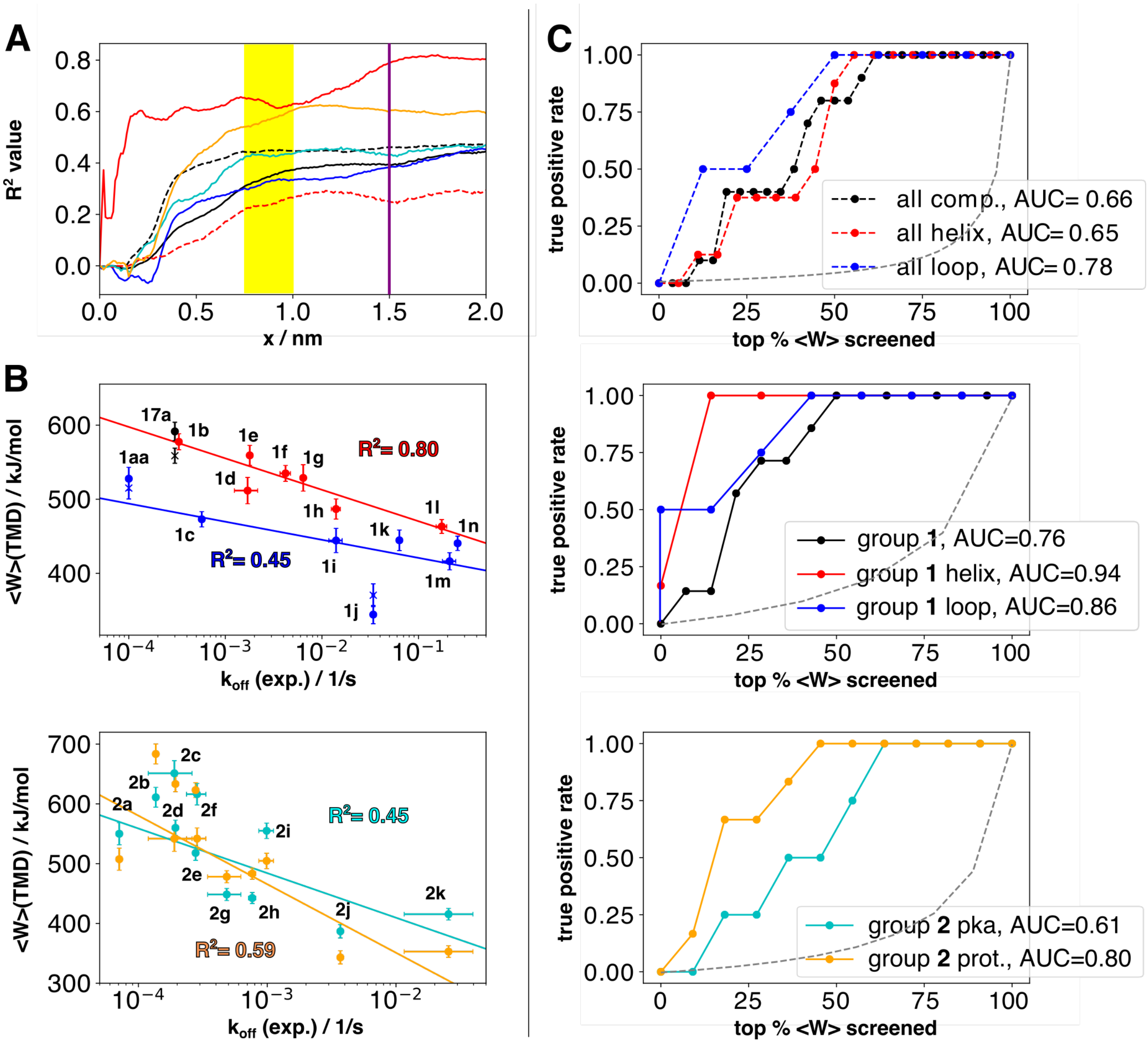

**Figure 2.** Comparison of experimentally derived $k_{off}$ constants and calculated TMD work <W>. Vertical error bars indicate the 1σ level from bootstrap analysis (see Methods), horizontal error bars indicate the standard error of the mean (SEM) for N=2-4 measurements. A: Pearsons' correlation coefficient $R^2$ as a function of pulling distance $x$. Transition state region highlighted in yellow, distance then the ligands have left the binding site in purple. Data on all compounds as dashed black lines, all helix compounds as dashed red lines, all loop compounds as dashed blue lines; group **1** compounds as black lines, group **1** helix binders as red lines, group **1** loop binders as blue lines; group **2** compounds in protonation states based on $pK_a$ predictions as cyan lines, protonated states as orange lines. B: <W> vs. $k_{off}$ plots at $x = 2.0$ nm. Group **1** helix binders in red, group **1** loop binders in blue, additional compound **17** in black. Protonation states chosen according to $pK_a$ prediction as dots, alternative protonation states as crosses. Group **2** protonation states chosen according to $pK_a$ prediction in cyan, alternative protonation states in orange. Linear regressions to Equation (6) as lines. C: true positive rate vs. top percent of <W> screened curves and area under curve (AUC) for <W>(TMD) as predictor for unbinding kinetics. Coloring according to A. Curve corresponding to random order displayed as dashed grey lines for optical reference.

We proceeded carrying out TMD simulations in strict non-equilibrium with the full compound groups **1** and **2**, with protonation states chosen according to $pK_a$ predictions (cf. Tables 1,2 and Figure 1 for an overview of all employed ligand structures), and used the resulting mean work <W> as unbinding scores for $k_{off}$.[28] Fitting theoretical and experimental results to Equation (8), Figure 2A displays Pearson's correlation coefficient $R^2$ for the full data set and different divisions into physiochemically relevant ligand groups resolved for the full range of pulling distances $x$. As can be seen, fitting the complete set of ligands leads to a maximal $R^2$ directly after the transition state region, which is in line with our assumption underlying Equation (7) that the non-equilibrium work is proportional to the transition state free energy. Interestingly, we observe that for group **1**, $R^2$ is maximal at 2 nm, and the later increase in $\mathbf{R^2}$ around 1.5–2.0 nm coincides with the presence of additional small barriers in the thermodynamic integration calculations (see Fig. S2A). We thus focus on an analysis of the given data set at $x = 2.0$ nm.

Fitting Equation (7) to the full data set on non-equilibrium works <W> as displayed in Fig. S5A, we yield a low $R^2 = 0.45$. It appears that for the full set of compounds, assuming $C$ in Eq. (5) to be constant is not a good approximation. We thus searched for physicochemically reasonable categories to group ligands according to comparable dissipative work. Based on differences in helix-ligand and loop-ligand contact dynamics,[70] we separated the compounds according to helix- and loop-binding compounds (see Fig. S5B), resulting in an improved $R^2 = 0.55$ for loop-binding compounds, but at an expense of $R^2 = 0.29$ for helix-binding compounds. We further separated the sets according to protein conformations into compound sets **1** (only taking resorcinol scaffolds into account) and **2** as displayed in Fig. 2B. In the case of group **1** compounds, this improved the $R^2 = 0.80$, but lead to a low $R^2 = 0.45$ for loop-binding compounds. Series **2** does not experience the split, as all contained compounds bind to the helix conformation. Fitting Equation (5) to this series however resulted only in a low $R^2$ = 0.45. While group **2** compounds are all predicted to be deprotonated, i.e., carry no net charge, we alternatively calculated the correlation coefficient for structures that carry a proton close to Asp93 as a test. Interestingly, this fit improved the agreement between experimental results and our theoretical model to a moderate $R^2$ = 0.59. Overall, it seems that a high <W> correlates with a small $k_{off}$.

To assess if <W>(TMD) is a suitable score for ranking ligands according to their $k_{off}$, we calculated receiver operating characteristic (ROC)-like curves in Fig. 2C characterized by the respective area under curve (AUC) for the given data set.[71,72] We plot the true positive rate against the top % according to the experimental $k_{off}$. Going through the data set from the highest to lowest $k_{off}$ value, we count a true positive if the current <W> value is the largest of all ligands that have not been screened so far. If <W> would be a perfect predictor, the AUC would be 1, and random order correspond to an AUC depending on the exact number of ligands in the investigated subgroup (AUC = 0.27 for N = 7 in group **1** loop binders and AUC = 0.13 for all N = 26 compounds, see grey dashed lines in Fig. 2C). While the application to both full and protein conformation-separated data set yielded only slight prediction power (AUC = 0.66-0.78), resorcinol compounds **1** after conformation separation resulted in a moderate to good prediction of slowly unbinding compounds (AUC = 0.76 for full group **1**, 0.86 for loop compounds and 0.94 for helix binding compounds). In this respect, <W>(TMD) faces similar problems with scaffold dependency like common affinity prediction-oriented docking,[73] but may indeed serve as a pre-selection criterion for slow unbinding compounds for suitable targets and ligands. In the case of compound group **2**, <W>(TMD) is a bad predictor (AUC = 0.61) for protonation states following $pK_a$ prediction, but becomes improved by using the alternative protonation states (AUC = 0.80). Based on this improvement of both $R^2$ and AUC, we tentatively formulate the hypothesis that the protonated states in group **2** which are unfavorable at pH = 7.5 might be involved in determining unbinding rates.

As all the calculations reported above took only unbinding along path 1 into account, we needed to assess if other possible unbinding pathways exist. Kokh et al. reported two unbinding routes for ligands that lead out of the binding site of the Hsp90 N-terminus,[20] the first one being path 1, and the second being found between helix 3 and the central β-sheet (path 2 in Fig. S1). Testing both pathways with **1a** and **2a**/**2aa**, we found that path 1 requires significantly less work for pushing the ligand into the solvent than path 2 (see Table S3). Furthermore, this pathway leads past Leu107, which has been implicated by point mutation experiments to affect unbinding kinetics.[7] We therefore judge path 1 to be the correct unbinding path, and path 2 to be irrelevant.

**Influence of protein conformation and electrostatics.** As a starting point for investigating molecular effects influencing unbinding rates, we focused on a dependence on helix/loop 3 conformation as implied by our analysis in Fig 2. In our simulations, helix binding compounds with decreasing $k_{off}$ display an increasing unbinding <W>. Based on experimental measurements and theoretical calculations it was proposed that entropic contributions from protein flexibility play a significant role in the determination of binding affinities for such compounds.[7] While we do not see a direct connection between <W> and entropic contributions in our simulations, the increase in non-equilibrium work thus might be connected with a decreased protein flexibility.

For loop-binding compounds, enthalpic contributions comprising electrostatic interactions were found to be key factors in determining $k_{off}$.[7] As can be seen in Fig. 2A and Table 2, the protonated ligand forms **1aa**, **17a** and **1ja** result in a slightly higher <W> than the neutral forms. This finding is consistent with the structure of the protein/ligand complexes, e.g., as the resulting ammonium moiety in **17a** is found close to Asp54 (see Fig. S6), allowing the formation of a salt bridge (N–O distances of 2.8 Å). In the case of compound **1aa**, <W> of **1a** is within the error range of **1aa** due to an increased distance of 4.7 Å between Asp54 and the protonable morpholine moiety in **1a** (Figure S6). This distance is sufficient to accommodate water molecules that shield a possible electrostatic interaction. In the case of **1ja**, the ligand sits within the binding crevice of the protein close to Asp93 (Figure S6). While the positive charge is located at the central pyrazole ring, the Asp93 side chain is within 5 Å without shielding by water molecules, resulting in a favorable electrostatic interaction. Apparently, within our investigated compound group **1**, ligand charge leads to higher <W>, and thus should result in slowing down unbinding kinetics in loop binding compounds.

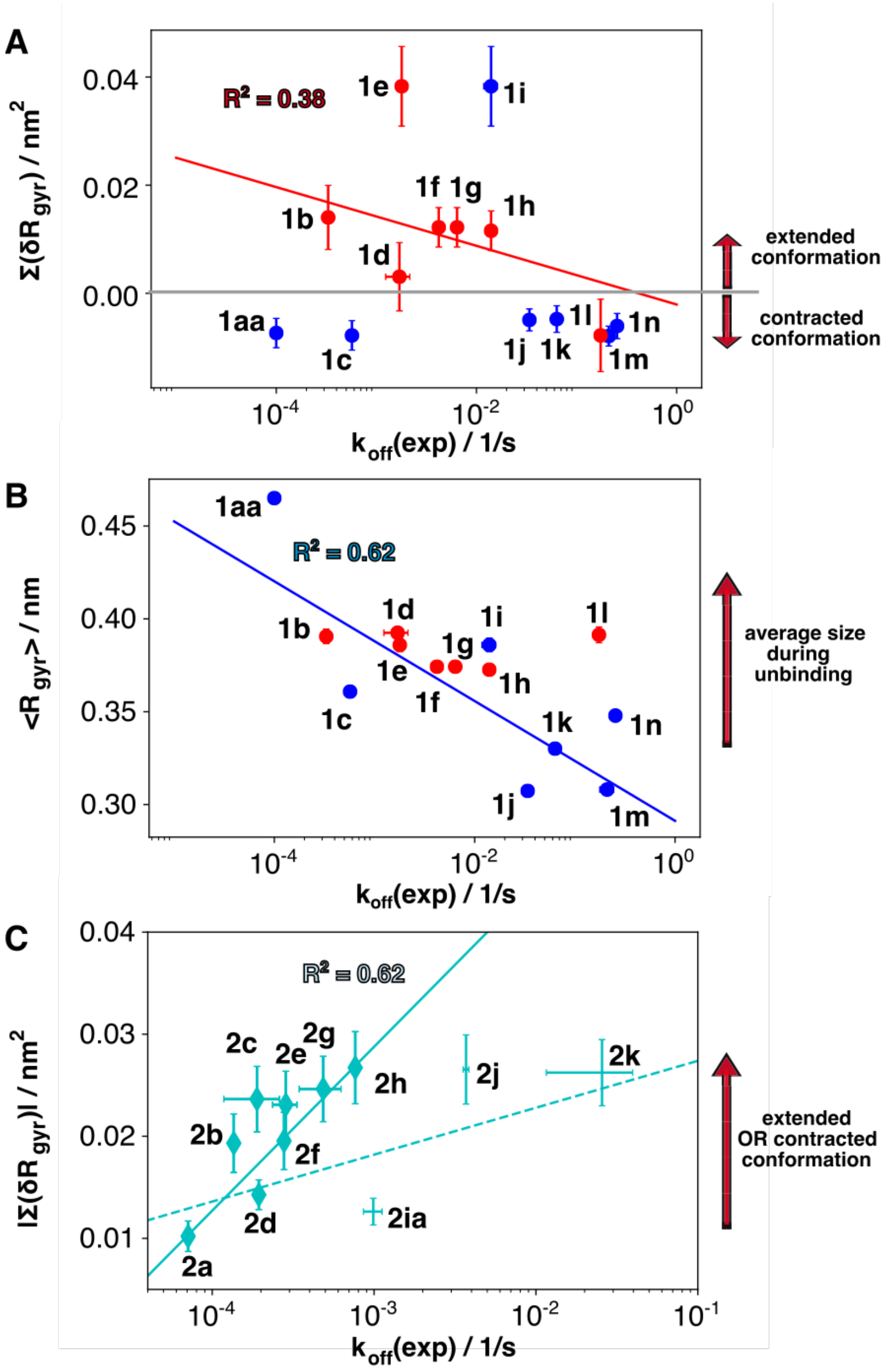

**Figure 3.** Influence of ligand radius on unbinding kinetics of the resorcinol series compounds. Group **1** helix binders in red, loop binders in blue, group **2** in cyan. Linear regression as full lines. Line colors match the color of data points used for fits. A: difference of radii of gyration in respect to the radius in the unbound state integrated over pulling distance. B: mean radius of gyration during unbinding. C: absolute integral of radii of gyration difference to unbound state over the full course of simulation in the N-heterocycle compound series **2**. Protonation states were chosen according to $pK_a$ predictions. Fit excluding outliers as full line, fit with outliers as dashed line.

Based on the predicted $pK_a$ values in Table 2, **1aa**, **17a** and **1j** should be the relevant structures at pH = 7.5. However, considering pH values at which the alternative protonation states claim 10% of the full ligand population (**1a**: 7.1; **17**: 8.8; **1ja**: 5.2), we realize that a small population of alternatively protonated forms could be present within an ensemble of protein-ligand complexes. Following our line of argumentation on path selection, these deprotonated ligand forms with smaller <W> may actually unbind faster than the protonated forms, and thus contribute to defining the ligand's unbinding rates.

Having investigated the impact of electrostatic interactions between protein and ligand on ligand unbinding kinetics, we followed up with an analysis of ligand conformational changes during unbinding. We note that the hypotheses listed in this and the next paragraph need to be taken with a grain of salt, as they are only weakly supported by our data, and in parts depend on a single data point (**1l** for helix binders and **1a** for loop binders). We used the ligand radius of gyration as observable, i.e., the average distance of all ligand atoms from their common center of mass, and compared the different radii with the natural logarithm of the experimentally determined $k_{off}$. For helix-binding compounds, we investigated a connection between the difference between the radius during unbinding and the average radius in the unbound state (i.e., during the 4$^{th}$ quarter of the simulation) integrated over the pulling distance with unbinding rates. This variable encodes if ligands bind to the protein in or need to pass through an extended conformation, and can be rationalized as an entropic contribution of the ligand itself to unbinding kinetics, i.e., if the conformational space of flexible ligands becomes restricted during unbinding. However, as can be seen in Fig. 3A, we only obtain a weak linear correlation of $R^2$ = 0.38.

In loop conformation binding compounds (Figure 3B), the overall radius of gyration (calculated for the unbound state) may decide the unbinding rate, though the agreement between linear fit and actual data is only moderate ($R^2$ = 0.62). Compound **1a** is significantly larger than the remaining loop binders. Loop-binding compounds appear therefore to unbind slowly if they exhibit strong van der Waals interactions with the protein, which again is in agreement with the importance of enthalpic contributions for loop binders.

Figure 3C shows that for the N-heterocycle series **2**, the best (though still moderate) agreement between radii of gyration changes and experimental unbinding constants for slowly unbinding compounds (i.e., when ignoring compounds **2i, 2j** and **2k**) is found for the absolute change in radius of gyration ($R^2 = 0.62$). Such outliers may be related to the large variation of side chains within the series, and **2j** and **2k** exhibit a unique scaffold. Within group **2**, slowly unbinding compounds appear to retain their shape during unbinding, which may be explained by a decreased molecular flexibility, while fast unbinding compounds can change their conformation, irrespective of if they pass through extended or contracted states. In summary, the detailed connection between conformational changes and unbinding kinetics for the investigated Hsp90 ligands appears to be non-trivial and highly dependent on the individual scaffold of a ligand.

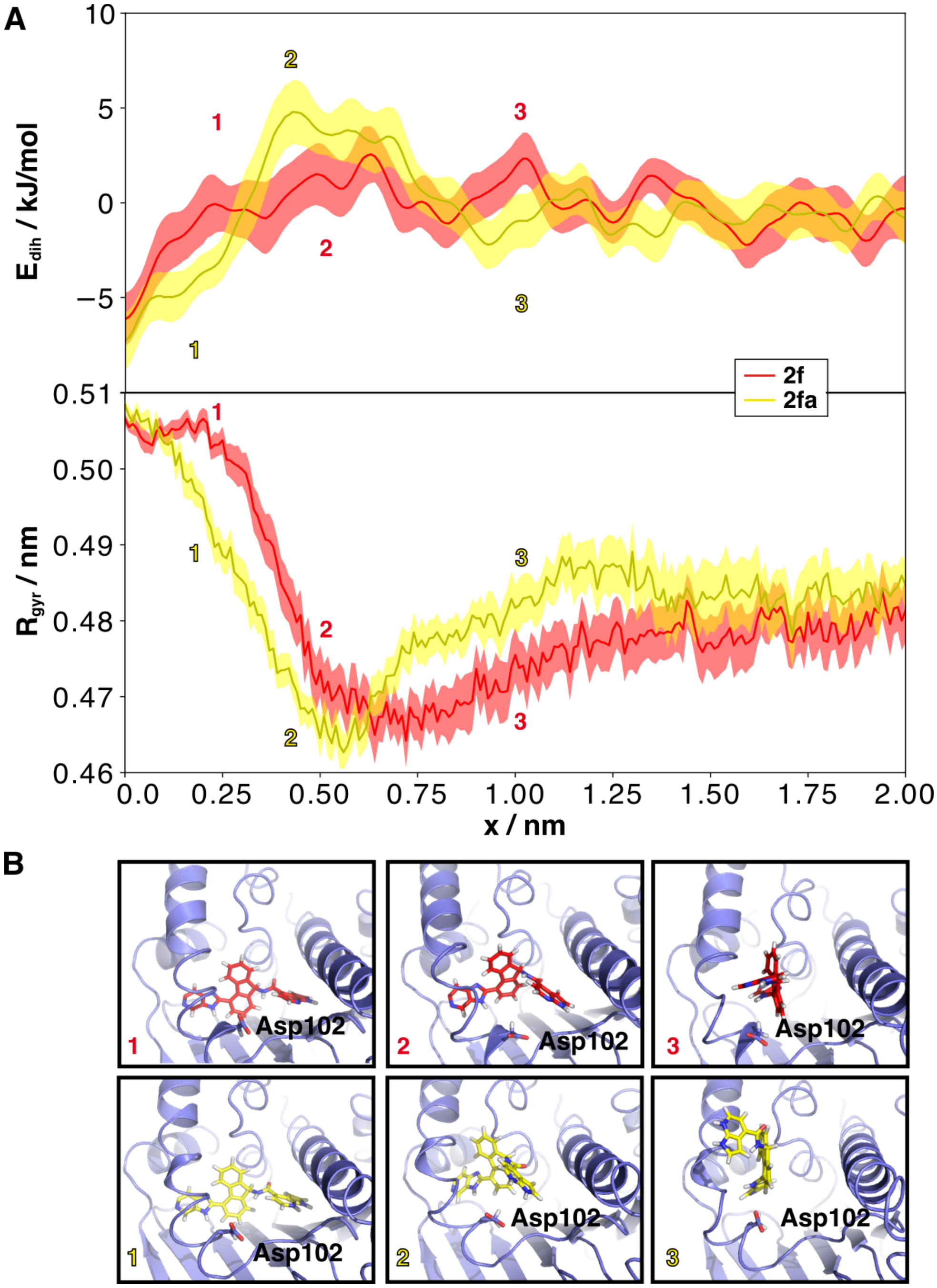

**Figure 4.** Electrostatic facilitation in compound **2f.** A: Ligand dihedral potential energies (referenced to the mean value of the last 0.5 nm and smoothed with a Gauss filter with σ = 0.2 nm) and radius of gyration as average of N = 30 simulations. Trajectory mean as lines, 1σ error level from bootstrap analysis (see Methods) as shaded area. Uncharged ligand **2f** in red, protonated form **2fa** in yellow. B: molecular details of ligand unbinding. A favorable charge interaction of the azaindoline moiety with Asp102 facilitates the contraction of the ligand from (1) to (2), and guides the ligand out of the binding pocket into the unbound conformation (3).

**Electrostatic locking vs. facilitation.** Focusing on the effect of electrostatics on <W> among group **2** compounds, we observe differences to series **1**. As can be seen in in Figure 2B and Table 2, compounds **2b**, **2d**, **2e**, **2g** and **2h** display an increased <W> for the protonated form like in the case of protonated group **1** compounds. Opposing this agreement, neutral ligands **2a**, **2c**, **2f** and **2i-k** exhibit a higher <W> than the protonated ligands with positive charge. However, no difference in binding positions can be observed, as their positive charge is placed at the same position as the pyrazole ring of **1ja** (for the example of compound **2aa**, see Figure S6). Figure 4 displays the detailed effect of molecular charges for the example of the unbinding pathway of **2f**: In the uncharged state, **2f** needs to stay in a more contracted conformation until a distance of 1.5 nm from the initial binding position. In the case of the protonated state **2fa**, a favorable charge interaction of the azaindole moiety with Asp102 occurs (see Figures 4B and S7), which allows a faster transition to the radius of gyration of the unbound state, which is already reached at a distance of 1.0 nm from the initial binding position, i.e., when the ligand is only partially unbound. As the ligand dihedral potential energy plot in Figure 4A shows, the contact with Asp102 allows **2fa** to access a conformation with higher energy during unbinding. Looking at the full group 2 radii of gyration and minimal distance to Asp102 time traces in Figure S7, we see that **2a**, **2c, 2f** and **2i** show a similar behavior, pointing to a common mechanism. We note that **2d** and **2e** exhibit radii and distance time traces that principally allow this described electrostatic facilitation of unbinding, as well, but deviate from the other ligands in details such as the comparable radius of gyration in bound and unbound state of **2d** and the surprising decrease in radius for **2ee**. Compounds **2j** and **2k** do not exhibit a contact with Asp102 during unbinding, and thus seem to follow a different mechanism of electrostatic-induced acceleration. The positive charge on group **2** ligands may thus facilitate the contraction of a ligand and guides it out of the binding pocket. This effect further offers an explanation for the improved $R^2$ and AUC of the fit of protonated group **2** ligands. As mentioned above, we propose that protonation state changes of individual ligands may appear transiently, so that the protonation state predicted by the linear regression of Equation (4) does not necessarily agree with the protonation state predicted from $pK_a$ calculations. While the $pK_a$ values of heterocycle side chains performing this contact are quite low (1.8-4.8), the 10% presence pH is found at 2.6-5.9. Especially in the case of the slowest unbinding compounds, fast protonation changes thus may act as a catalyst for the acceleration of unbinding instead of locking the ligand to the protein. We note however that we observe no correlation between the predicted $pK_a$ values and the $k_{off}$ of compounds. In any way, from a drug design perspective, it is interesting to notice that fixed charges on a ligand do not necessarily increase residence times, but may facilitate unbinding by transient electrostatic interactions with the protein.

**Performance and applicability of non-equilibrium TMD.** In recent years, several other novel methods have been established for fast and efficient computation of binding kinetics[12,20,66,70-72] (see refs. 73,74 for reviews), and our approach presented here shares similarities with methods based on metadynamics[12] and steered MD.[74] While the calculation of a constraint numerically is more complex than the addition of an additional bias potential, the major advantage of our method is that employing a constraint allows to scan the preset pulling coordinate with a linear velocity that is accurate down to machine precision. Furthermore, we experienced in other works[63] that employing biasing potentials causes problems at steep potential gradients, causing long simulation times or even simulation crashes. Such problems are overcome by employing constraints, as the biasing force $f_c$ exactly cancels out such gradients. As a prerequisite, we need to have initial information on unbinding pathways to create a suitable reaction coordinate to apply the target bias: as stated in the Introduction, while the ligand is free to diffuse perpendicular to the pulling vector, this dynamics is restricted by the ratio between diffusion rate and constraint velocity. At the start of the simulation, the initial pulling vector thus needs to roughly point into the direction of the presumed binding pathway. Besides taking educated guesses, this information can be provided by other methods[20,66,75] that employ more general reaction coordinates. It was recently shown that TMD simulations can be used to effectively push a molecular system of choice along a reaction coordinate that correctly mimics the pathway taken under equilibrium conditions.[28] The first major strength of our non-equilibrium method is the significant reduction of necessary computational power: unbinding can be enforced within 0.1 to 0.3 ns simulated time, which allows us to reduce the necessary calculation time by a factor of 30 in comparison to stationary TI[75] (180 CPUhs for one non-equilibrium unbinding event sampled with 30 trajectories vs. 6000 CPUhs for one equilibrium free energy pathway analyzed by 20 intermediate steps on a recent octacore CPU workstation). Secondly, the non-equilibrum work rapidly converges (see Figure S4),[29] and each simulation by definition results in an unbinding event, which reduces the necessary number of simulations to a number well below that for Markov State Model creation.[76] Thirdly, we do not change the full system Hamiltonian, but merely add a perturbation, avoiding artifacts such as protein unfolding that appear in smoothed/scaled MD.[77,78]

Summarizing the results of our investigation, it becomes clear that estimation of binding / unbinding rates and elucidating the relevant underlying molecular mechanisms is far more complex than free energy calculations of binding poses, as it not only require to correctly characterize a single binding pose, but to assess the full dynamics along binding/unbinding pathways. While binding simulations apparently are more easy to carry out due to the shorter inherent timescales, special care needs to be taken that the simulations result in the correct binding pose.[30] In this respect, biased unbinding simulations are simpler, as they only require enforced unbinding, and it is always "simpler to break things than to make things". However, to gain meaningful results that can be compared to experimental data, the biased simulations need to reproduce the correct protein-ligand dynamics, i.e., ligands need to leave the protein along the correct pathway, which needs to be identified first. Another challenge is found in the theoretical basis of the Jarzynski equality (4) itself: biased simulations need to start from a representative equilibrium ensemble of the employed protein. In our case, we were interested in developing a “fast” method for prediction of unbinding kinetics, and thus performed only short equilibration runs of 0.1 ns before initiating unbinding. This requires that the employed protein crystal structure is close to an equilibrium structure. Furthermore, if unbinding kinetics are coupled to protein conformational dynamics, e.g., by only shortly opening a binding site to allow unbinding, our approach will not succeed. Besides the results presented here, we found that flexible proteins with challenging unbinding pathways pose a problem for our non-equilibrium TMD method, as a similar investigation with ligands bound to the $\beta_2$ adrenergic receptor[79] performed by us did not succeed in obtaining a successful linear non-equilibrium energy relationship (data not shown). The reason for this appears to be the presence of a second intermediate ligand binding site[80] that increases the complexity of the underlying ligand diffusion pathways, and protein conformational dynamics. Last, as the details of ligand conformational dynamics and protein-ligand charge interactions during unbinding seem to be crucial for correct predictions, small errors in ligand parameterization can cause significant problems, which require elaborate parameterization schemes. As we want to be able to perform calculations with a larger number of ligands, we here use a quick standard parameterization scheme via antechamber[44] using semi-empirical charges. It thus may be that a part of the spread of data points around linear fits in Fig. 2B is a result from errors in dihedral angle potentials and atomic charge assignment. In RAMD simulations on Hsp90 it was indeed shown that including charge details like halogen $\sigma$-holes improves the prediction of unbinding kinetics.[20] As the problems listed above are principal effects coming from the underlying Hamiltonian dynamics of the protein-ligand complexes, other biased simulations approaches[78] will face similar challenges.

## Conclusion and future perspective

To elucidate the molecular determinants for unbinding kinetics, we here combined preexisting and novel data from SPR binding kinetics measurements and X-ray crystallography with non-equilibrium targeted MD simulations on the N-terminal domain of Hsp90 for two compound series. The non-equilibrium work <W> obtained from TMD simulations converges quickly, and is a promising predictor for slowly unbinding compounds. The extraction of clear-cut molecular discriminators determining unbinding kinetics however proved to be difficult, and yielded only moderate correlations. Apparently, the connection between protein-ligand interactions, ligand conformational changes and unbinding rates is complex, and requires further research. We propose ligand conformational changes and nonbonded protein-ligand interactions to have an impact on unbinding rates. The exact impact of these two contributions apparently depends on individual ligand scaffolds and the details of transient protein-ligand interaction during unbinding. Electrostatics may exhibit a dual effect onto unbinding kinetics: the presence of a charge can either increase the residence time of compounds by locking it to the protein, or accelerate unbinding by facilitating the formation of a contracted form and guiding the ligand out of the binding pocket. Concerning consequences for rational drug design in general, we thus propose that a clear knowledge of the conformational space accessible by the ligand, the exact unbinding pathway, and the transient protein-ligand interactions along this path are requirements for the prediction of ligands with favorable unbinding kinetics.

As our interpretation of the mean non-equilibrium work <W> as score for $k_{off}$ by use of Equation (4) is based on the Jarzynski equality,[65] we potentially can calculated the unbinding free energy profile directly from <W>. Indeed, we recently showed for a NaCl/water test system that such a correction can readily be achieved via dissipation-corrected targeted MD simulations.[29] As this approach additionally yields friction profiles, we will aim to use the resulting information to carry our Langevin Dynamics calculations[81] for the prediction of absolute ligand unbinding kinetics.

## Associated Content

## Supporting Information

Three supporting tables, seven supporting figures (PDF)

SMILES annotations (CSV)

## Accession Codes

The crystallographic coordinates of novel compound **2d** are deposited in the Protein Data Bank under the accession code 5LRL. Authors will release the atomic coordinates and experimental data upon article publication.

## Author Information

## Corresponding Author Information

*phone: +49-761-203-5913

eMail: steffen.wolf@physik.uni-freiburg.de

## Author contributions

[#]S.W. and M.A. contributed equally to this work. S.W. designed and performed all computational studies. M.A. designed and performed SPR and X-ray studies. M.L., F.V. and D.M. solved selected crystal structures. J.G. assisted in data analysis. J.B., M.K.D. M.F., J.S. and KG supervised the studies. All authors contributed to writing of the manuscript.

## Acknowledgements

This work was supported by the EU/EFPIA Innovative Medicines Initiative (IMI) Joint Undertaking K4DD (grant No. 115366). This paper reflects only the authors' views and neither the IMI nor the European Commission is liable for any use that may be made of the information contained herein. We acknowledge the Partnership for Advanced Computing in Europe for awarding us access at CINECA Italy (project No. 2015133089). We are grateful to M. Bianciotto, D. Kohk and R. Wade for helpful discussions.

**Abbrevations Used**

Hsp90: heat shock protein 90; MD: molecular dynamics; SPR: surface plasmon resonance; TMD: targeted molecular dynamics